\documentclass[aps,prc,superscriptaddress,reprint,amsmath]{revtex4-1}

\usepackage{graphicx,dcolumn}
\newcommand\etal{\textit{~et~al.}}\makeatletter

\begin{document}

\title{Symmetry energies for $A = 24$ and 48 and the USD and KB3 shell
  model Hamiltonians}

\author{A. Kingan}
\affiliation{Department of Physics and Astronomy, Rutgers University,
  Piscataway, New Jersey 08854, USA}

\author{K. Neerg\aa rd}
\affiliation{Fjordtoften 17, 4700 N\ae stved, Denmark}

\author{L. Zamick}
\affiliation{Department of Physics and Astronomy, Rutgers University,
  Piscataway, New Jersey 08854, USA}

\begin{abstract}

  Calculations in the \textit{sd} and \textit{pf} shells reported some
  time ago by Satu\l a\etal\ [Phys.~Lett.~B~407, 103 (1997)] are
  redone for an extended analysis of the results. As in the original
  work, we do calculations for one mass number in each shell and
  consider in each case the sequence of lowest energies for isospins
  0, 2, and 4, briefly the symmetry spectrum. Following further the
  original work we study how this spectrum changes when parts of the
  two-nucleon interaction are turned off. The variation of its width
  is explored in detail. A differential combination
  $\epsilon_\text{W}$ of the three energies was taken in the original
  work as a measure of the so-called Wigner term in semi-empirical
  mass formulas, and it was found to decrease drastically when the
  two-nucleon interaction in the channel of zero isospin is turned
  off. Our analysis shows that the width of the symmetry spectrum
  experiences an equally drastic decrease, which can be explained
  qualitatively in terms of schematic approximations. We therefore
  suggest that the decrease of $\epsilon_\text{W}$ be seen mainly as a
  side effect of a narrowing of the symmetry spectrum rather than an
  independent manifestation of the two-nucleon interaction in the
  channel of zero isospin.

\end{abstract}

\maketitle

\section{\label{sec:in}Introduction}

The article~\cite{ref:Sa97} by Satu\l a\etal\ was pivotal in
discussions around the turn of millennium of the roles of different
parts of the effective two-nucleon interaction in the formation of
patterns of binding energies of nuclei with approximately equally many
neutrons and protons. The authors calculate ground state energies in
the \textit{sd} and \textit{pf} shells from the isobarically invariant
Hamiltonians USD~\cite{ref:Wi82} and KB3~\cite{ref:Po81},
respectively. For the analysis of their results they use two different
differential combinations of the ground state energies of doubly even
nuclei with pairs of the neutron number $N$ and the proton number $Z$
in the vicinity of a given pair with $N = Z$. One combination involves
three nuclei with the same $A = N + Z$, and one involves eight nuclei
spanning five different $A$. Our present study focuses on the first
type of analysis, which displays the pattern of ground state energies
along a single isobaric chain.

For a given $A$ we denote by $E(T)$ the calculated lowest energy of a
state with isospin $T$, which is also the ground state energy when $N$
and $Z$ are even and $|N - Z|/2 = T$. Two parameters
$\epsilon_\text{W}$ and $\kappa$ are defined by
\begin{equation}\label{eq:epWkap}
  E(T) = E(0) + \epsilon_\text{W} + \tfrac12 \kappa T^2 \quad
  \text{for $T = 2$ and 4} . 
\end{equation}
Satu\l a\etal\ consider the parameter $\epsilon_\text{W}$, which they
take as a measure of the so-called Wigner term included by Myers and
Swiatecki in their semi-empirical mass formula~\cite{ref:My66}. For $A
= 24$ and 48 they calculate $\epsilon_\text{W}$ both for the full
Hamiltonian and for variants where components of the two-nucleon
interaction in the isospin $T' = 0$ channel are turned off
successively. (We denote in this article the two-nucleon isospin by
$T'$ to distinguish it from the isospin $T$ of the nucleus.) They find
that when the entire $T' = 0$ interaction is turned off,
$\epsilon_\text{W}$ vanishes completely or almost completely. They
conclude that the $T' = 0$ interaction is responsible for the Wigner
term.

An analysis of the sequence of energies $E(T), \linebreak T = 0, 2,
4$, briefly the symmetry spectrum, in terms of the single parameter
$\epsilon_\text{W}$ is incomplete. To complete the analysis and
examine in particular the behavior of the parameter $\kappa$ in
Eq.~\eqref{eq:epWkap} we redid the calculations of
Ref.~\cite{ref:Sa97}. The results are presented in Secs.~\ref{sec:sym}
and \ref{sec:eW}. In Sec.~\ref{sec:sym} the dependence of $E(4) -
E(0) = \epsilon_\text{W} + 8 \kappa$ on the set of components included
in the interaction is explored. It is found to vary as drastically as
$\epsilon_\text{W}$. Most of this varation is shown to have
straighforward explanations in terms of schematic approximations. The
parameter $\epsilon_\text{W}$ is then discussed in Secs.~\ref{sec:eW}.
Its variation follows mostly that of $E(4) - E(0)$, which leads to the
suggestion that the reduction of $\epsilon_\text{W}$ observed in
Ref.~\cite{ref:Sa97} be seen mainly as a side effect of a narrowing of
the symmetry spectrum rather than an independent manifestation of the
$T' = 0$ interaction.

Previously, Bentley\etal\ analyzed along the same lines results from a
calculation for $A = 48$ using the $1f_{7/2}$ valence
space~\cite{ref:Be14}. In Sec.~\ref{sec:oth} we briefly summarize
their results and present them in a form compatible with the
presentation of our present ones. They are found mainly representative
for the KB3 results, which corroborates that the $1f_{7/2}$ shell
seems to dominate these results. Section~\ref{sec:con} concludes the
article.

\section{\label{sec:sym}Width of the symmetry spectrum}

\begin{figure}
  \includegraphics{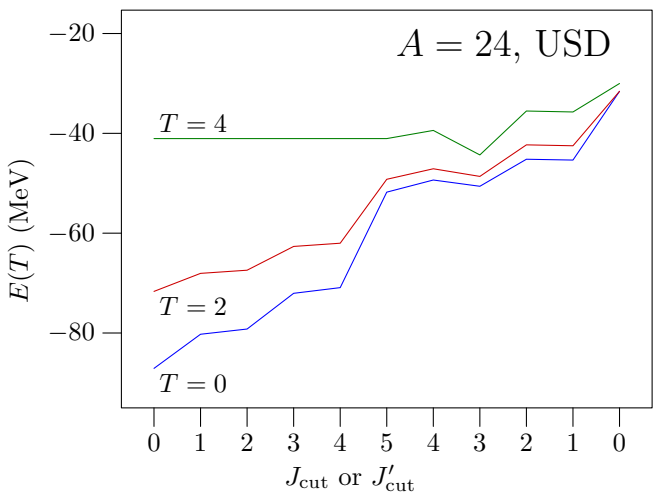}
  \caption{\label{fig:eusd}(Color online) The energy $E(T)$ for
    \mbox{$A = 24$} and $T = 0$, 2, and 4 calculated from the
    Hamiltonian USD as a function of the angular momenta
    $J_\text{cut}$ and $J_\text{cut}'$. The increasing sequence of
    abscissa in the first part of the axis are $J_\text{cut}$, and the
    subsequent decreasing sequence  $J_\text{cut}'$.}
\end{figure}

\begin{figure}
  \includegraphics{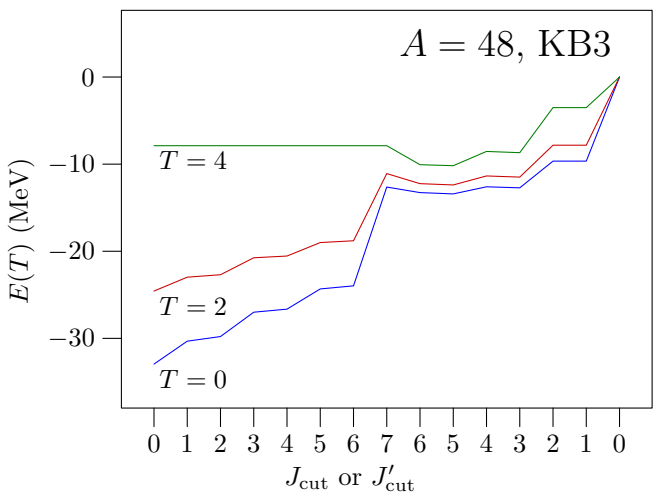}
  \caption{\label{fig:ekb3}(Color online) As Fig.~\ref{fig:eusd} for
    $A = 48$ and the Hamiltonian KB3.}
\end{figure}

Calculations for $A = 24$ with the Hamiltonian USD and $A = 48$ with
the Hamiltonian KB3 were done with the program
\textsc{nushellX@msu}~\cite{ref:Br14} for $T = 0$, 2, and 4. In each
case the lowest energy was calculated from the full Hamiltonian and
with components of the two-nucleon interaction successively turned
off. First the components with $T' = 0$ and
$J = 1, 2, \dots, J_\text{max}$ were turned off in this order, and
then the components with $T' = 1$ and $J = J_\text{max}-1,
J_\text{max}-2, \dots, 1$ in this order, where $J_\text{max}$ is the
maximal two-nucleon angular momentum of the valence space. In the
\textit{sd} space, $J_\text{max} = 5$, and in the \textit{pf} space,
$J_\text{max} = 7$. Note that in these valence spaces, no pair has
$T' = 0$ and $J = 0$, and no pair has $T' = 1$ and $J = J_\text{max}$.
Each Hamiltonian is thus characterized by angular momenta
$J_\text{cut}$ and $J_\text{cut}'$. The $T' = 0$ interaction matrix
elements are zero for $J \le J_\text{cut}$, and the\linebreak
$T' = 1$ matrix elements for $J \ge J_\text{cut}'$. Because of the
order in which the components are turned off, either $J_\text{cut}$ or
$J'_\text{cut}$ is always equal to $J_\text{max}$. Therefore, only one
of them needs mention to characterize the Hamiltonian. Thus, for
example, $J_\text{cut} = 3$ means that all $T' = 0$ matrix elements
with $1 \le J \le 3$ are set to zero, and $J'_\text{cut} = 2$ means
that all $T' = 0$ matrix elements and all $T' = 1$ matrix elements
with $2 \le J \le J_\text{max} - 1$ are set to zero. In particular,
$J_\text{cut} = 0$ indicates the full Hamiltonian, and
$J_\text{cut}' = 0$ is the Hamiltonian of non-interacting nucleons.
Both $J_\text{cut} = J_\text{max}$ and $J_\text{cut}' = J_\text{max}$
mean that the entire $T' = 0$ interaction is turned off, so when this
case is referred to in our figures and tables, the label may be
understood as a value of $J_\text{cut}$ or a value of $J'_\text{cut}$;
this makes no difference. In the calculations, first $J_\text{cut}$
ascends from 0 to $J_\text{max}$, when it equals $J_\text{cut}'$;
thence $J_\text{cut}'$ descends to 0.

The calculated energies $E(T)$ are shown in Figs.~\ref{fig:eusd} and
\ref{fig:ekb3} as functions of $J_\text{cut}$ or $J_\text{cut}'$. The
energy $E(4)$ is constant as long as only $T' = 0$ interactions are
turned off. This is because to make $T = 4$ with eight valence
nucleons, every pair must have $T' = 1$, so these states do not feel
the $T' = 0$ interaction. In every case of $A$, $J_\text{cut}$, and
$J_\text{cut}'$, the energy $E(T)$ increases with $T$. The difference
$E(T) - E(0)$ is the symmetry energy. For $A = 48$ the symmetry
spectrum shrinks to degeneracy in the absence of interactions because
the system of eight $1f_{7/2}$ nucleons can have $T = 4$. For $A = 24$,
one has in this case \linebreak $E(0) = E(2)$ while $E(4) - E(2)$
equals twice the spacing of the $1d_{5/2}$ and $2s_{1/2}$ levels.

\subsection{\label{sec:48}$A = 48$}

We begin our discussion with Fig.~\ref{fig:ekb3} because the case
$A = 24$ is slightly more complicated. First notice that the symmetry
spectrum almost does not change when components of the interaction
with even $T' + J$ are turned off. This is a natural consequence of
the absence of pairs with these quantum numbers from the $1f_{7/2}$
shell. Thus, in a calculation of the energy by perturbation theory
starting from a configuration in the $1f_{7/2}$ shell, terms involving
the said components would not appear before the third order in the
interaction.

\subsubsection{\label{sec:m}Monopole interaction}

One can attempt to simulate the variation of the symmetry spectrum by
schematic interactions. First assume that the two-nucleon interaction
$v$ depends on nothing but $T'$, that is,
\begin{equation}\label{eq:T'only}
  v = v_\text{m} =  - \sum_{T'} G_{T' } P_{T'} ,
\end{equation}
where $G_{T'}$ are constants, and $P_{T'}$ projects on isospin $T'$.
Using
\begin{equation}\label{eq:proj}
  P_{T'} = \begin{cases}
    \tfrac14 - \boldsymbol{t}_1 \cdot \boldsymbol{t}_2 , & T' = 0 , \\
    \tfrac34 + \boldsymbol{t}_1 \cdot \boldsymbol{t}_2 , & T' = 1 ,
  \end{cases}
\end{equation}
where $\boldsymbol{t}_i$ is the nucleonic isospin, one gets by an easy
calculation (given, for example, in Ref~\cite{ref:Sh63}) that in a
state with isospin $T$, the interaction term in the Hamiltonian has
expectation value
\begin{multline}\label{eq:Em}
  E_\text{int} = - \frac n 8 [ (n + 2) G_0 + 3 (n - 2) G_1 ] \\
      + \tfrac12 ( G_0 - G_1 ) T(T + 1)
    \mathrel{\mathop:}= - \sum_{T'} G_{T' } n_{T'}  ,
\end{multline}
where $n = 8$ is the number of valence nucleons. The factors $n_{T'}$
in the last expression are the numbers of bonds with isospin $T'$ in
the system with $n$ nucleons and isospin $T$. In any state of this
system, the two-nucleon density matrix $\rho_2$ satisfies
\begin{equation}
  \text{tr\,} \rho_2 P_{T'} = n_{T'}
\end{equation}
and
\begin{equation}
  E_\text{int} = \text{tr\,} \rho_2 v .
\end{equation}
It follows that Eq.~\eqref{eq:Em} is exact for lowest-energy states
with isospin $T$ provided
\begin{equation}\label{eq:rho-centr}
  -G_{T'} = \frac { \text{tr\,} \rho_2(T) P_{T'} v }
  { \text{tr\,} \rho_2(T) P_{T'} } ,
\end{equation}
where $\rho_2(T)$ is the two-nucleon density matrix in one such state.
By the charge independence of $v$, the centroid \eqref{eq:rho-centr}
does not depend on which particular substate of the lowest-energy
isospin multiplet generates $\rho_2(T)$.

The expectation value of the total Hamiltonian is
\begin{equation}\label{eq:tot}
  E = \text{tr\,} \rho_1 h + E_\text{int} ,
\end{equation}
where $\rho_1$ and $h$ are the density matrix and Hamiltonian of a
single nucleon. Provided the first term in this expression is
constant, the symmetry spectrum is thus given by this constant plus
the expression \eqref{eq:Em} in so far as the
centroids \eqref{eq:rho-centr} do not depend on $T$. From a deviation
of the symmetry spectrum from this form, and from the $T(T + 1)$
proportionality of the symmetry energy, in particular, on can infer,
conversely, that either the variation of $T$ redistributes the nucleon
on the single-nucleon levels, or the centroid \eqref{eq:rho-centr}
depends on $T$, or both.

The interaction \eqref{eq:T'only} is seen from Eq.~\eqref{eq:proj},
to be a monopole interaction. It was shown by
Caurier\etal~\cite{ref:Ca05} that the monopole part of the interaction
in the pair of orbits with angular momenta $j_1$ and $j_2$ is given by
\begin{equation}\label{eq:p}
  -G_{T'} = \frac { \sum_J (2J + 1) V_{T'J}(j_1j_2j_1j_2) }
    { \sum_J (2J + 1) } ,
\end{equation}
where $V_{T'J}(j_1j_2j'_1j'_2)$ is the interaction matrix element in
the notation of, for example, Eq.~(17.46) of Ref.~\cite{ref:Ta93}. The
summations in Eq.~\eqref{eq:p} run over the $J$ that are compatible
with $j_1$, $j_2$, and $T'$, and the weight $2J + 1$ is proportional
to the dimension of the two-nucleon space with quantum numbers $j_1$,
$j_2$, $T'$, and $J$. Centroids covering a part of the valence space
can also be defined,
\begin{equation}\label{eq:set}
  -G_{T'} = \frac { \sum_{j_1,j_2 \in \mathcal S,j_1 \le j_2,J}
    (2J + 1) V_{T'J}(j_1j_2j_1j_2) }
    { \sum_{j_1,j_2 \in \mathcal S,j_1 \le j_2,J} (2J + 1) } ,
\end{equation}
where $\mathcal S$ is some set of orbits. In the configuration of the
fully occupied space $\mathcal S$ (which has $T = 0$), the two-nucleon
density matrix $\rho_2$ is the unit matrix. The centroids
\eqref{eq:rho-centr} and \eqref{eq:set} then coincide, so one gets the
interaction energy $E_\text{int}$ in this configuration by inserting
the expressions \eqref{eq:set} into the expression \eqref{eq:Em}. We
shall explore what may be learned from assuming identity of these
centroids also in the present case of an open $1f_{7/2}$ shell.

\begin{table}
  \caption{\label{tbl:kb3}Results in the monopole
    approximation for $A = 48$ and the KB3
    Hamiltonian. The set $\mathcal S$ in
    Eq.~\eqref{eq:set}  is
    indicated in the first headline, and the case of $J_\text{cut}$ or
    $J'_\text{cut}$ is described in the first column in the same way
    as in the figures: The initial increasing sequence of integers are
    $J_\text{cut}$, and the subsequent decreasing sequence
    $J'_\text{cut}$. Results for the $1f_{7/2}$ subspace are shown only
    when different from the preceding row, that is, when the $1f_{7/2}$
    shell allows pairs with $T' = 0$ and $J = J_\text{cut}$ or $T'
    = 1$ and $J = J'_\text{cut}$. All energies are in MeV.}
\begin{ruledtabular}
\begin{tabular}{ddddd}
\multicolumn{1}{c}{$J_\text{cut}$ or}&
\multicolumn{2}{c}{$1f_{7/2}$}&
\multicolumn{2}{c}{\textit{pf} shell}\\
\multicolumn{1}{c}{$J'_\text{cut}$}&
\multicolumn{1}{c}{$E_\text{m}(0)$}&
\multicolumn{1}{c}{$G_0\!-\!G_1$}&
\multicolumn{1}{c}{$E_\text{m}(0)$}&
\multicolumn{1}{c}{$G_0\!-\!G_1$}\\\hline
0&-19.36&1.26&-14.01&1.41\\
1&-18.38&1.16&-12.40&1.25\\
2&&&-10.77&1.08\\
3&-16.71&1.00&-7.92&0.80\\
4&&&-6.03&0.61\\
5&-14.54&0.78&-3.27&0.33\\
6&&&-1.71&0.18\\
7&-4.33&-0.24&0.04&0.00\\
6&-5.84&-0.32&0.55&0.03\\
5&&&-0.39&-0.02\\
4&-4.74&-0.26&-0.79&-0.04\\
3&&&-1.62&-0.09\\
2&-1.23&-0.07&-0.36&-0.02\\
1&&&-0.40&-0.02
\end{tabular}
\end{ruledtabular}
\end{table}

We now describe the scheme of calculation employed for this purpose in
its general form because it will be used also in the case of $A = 24$.
The energy $E(T)$ is calculated by Eq.~\eqref{eq:tot}, where
$\text{tr\,}\rho_1 h$ is taken to be constant. For the interaction
energy $E_\text{int}$ we use the expression \eqref{eq:Em} with
$-G_{T'}$ given by Eq.~\eqref{eq:set}. We denote the energy $E(T)$
obtained in this approximation, which we call the \textit{monopole}
approximation, by $E_\text{m}(T)$ to distinguish it from the energy
given by the full Hamiltonian. For $A = 48$, we take
$\text{tr\,}\rho_1 h = 0$ because the $1f_{7/2}$ orbit has zero energy
in the KB3 Hamiltonian. The results for two different $\mathcal S$ are
shown in Table~\ref{tbl:kb3}. One set consists of only the orbit
$1f_{7/2}$, and one set includes the entire \textit{pf} shell. Shown
in the table are $E_\text{m}(0)$ and $G_0 - G_1$, which is the
coefficient of $\tfrac12 T(T + 1)$ in Eq.~\eqref{eq:Em}.

The calculation with $1f_{7/2}$ centroids is seen to reproduce quite
well the energies $E(0)$ in Fig.~\ref{fig:ekb3} except for a general
upwards shift by roughly 10~MeV. The difference $E_\text{m}(4) - E(4)$
is about 1~MeV almost independently of $J_\text{cut}$ and
$J'_\text{cut}$. Necessarily, it does not depend on $J_\text{cut}$.
Thus, roughly, $E_\text{m}(T) - E(T)$ depends on $T$ alone for $T = 0$
and 4. More precisely, $E_\text{m}(0) - E(0) = 13.7$~MeV for
$J_\text{cut} = 0$, and 8.3~MeV for $J'_\text{cut} = 1$. This decrease
may be understood by observing that in the diagonalization, the system
seeks to utilize the most attractive part of the interaction rather
than its average, and the full Hamiltonian offers more opportunities
for this optimization than the Hamiltonian with only the $J = 0$
interaction.

Even the pattern of changes of $E(0)$, including, in particular, a
decrease when the $T' =1, J = 6$ interaction is turned off, and the
approximate size of each shift, is reproduced in detail. It follows
from Eqs.~\eqref{eq:tot}, \eqref{eq:Em}, and \eqref{eq:p} that each
shift of $E_\text{m}(0)$ is proportional to only one interaction
matrix element, namely the $1f_{7/2}$ shell matrix element with the
$T'$ and $J$ that are turned off. The coefficient of proportionality
is negative and depends on $T'$ and $J$.

The calculation with \textit{pf} shell centroids is much less
successful. The general slope of increase of $E_\text{m}(0)$ is
further below that of $E(0)$ than in the $1f_{7/2}$ case, and, at
variance with the KB3 results, fairly large shifts occur when
components of the interaction that are not active in the $1f_{7/2}$
shell are turned off. The shift when the $T' = 0, J = 7$ interaction
is turned off is much smaller than in the KB3 results and the
calculation with $1f_{7/2}$ centroids. The deviation from the latter
is the natural result of the $1f_{7/2}$ matrix element being the only
one with these quantum numbers. This single matrix element indeed has
a much smaller relative weight in the \textit{pf} shell centroid
$-G_0$ than in the corresponding $1f_{7/2}$ centroid. Our analysis
thus suggests a dominance of the $1f_{7/2}$ shell in the KB3 results.

\subsubsection{\label{sec:p}Pairing}

As is evident from Fig.~\ref{fig:ekb3} and Table~\ref{tbl:kb3}, even
with $1f_{7/2}$ centroids, the monopole approximation fails badly in
more than one respects to reproduce the KB3 results. We mentioned
already the discrepancy $E_\text{m}(0) - E(0) \approx 10$~MeV.
Moreover, the monopole approximation gives \emph{decreasing} symmetry
spectra after all $T' = 0$ interactions have been turned off. This is,
in fact, a direct consequence of $E_\text{int}(0)$ and $G_0 - G_1$
having equal signs by the expression \eqref{eq:Em} when $G_0 = 0$. In
particular $E_\text{int}(0) < 0$ implies $G_0 - G_1 < 0$. A path to
simultaneous remedy of both failures is suggested by the observation
that the $J = 0$ part of the KB3 interaction is similar to the pairing
interaction $v_\text{p}$, which, in a single oscillator shell, has
matrix elements
\begin{equation}\label{eq:vp}
  V_{T'J}(j_1j_2j'_1j'_2) =
    - \tfrac12 \delta_{J 0} G \sqrt{ (2 j_1 + 1) (2 j_1' + 1) }
\end{equation}
with a constant $G$. (Note that  in a single oscillator shell,
$\delta_{J 0}$ vanishes unless $j_1 = j_2$, $j_1' = j_2'$, and
$T' = 1$.) Indeed every KB3 matrix element $V_{10}(jjj'j')$ is
negative, and the ratio of the largest and the smallest $G$ extracted
from these matrix elements by means of Eq.~\eqref{eq:vp} is 4.6. The
value $G = 0.48$~MeV extracted from the $1f_{7/2}$ matrix element is
close to both the arithmetic and the geometric mean.

For $G > 0$ the lowest eigenvalue of $v_\text{p}$ with a given $T$ is
\begin{equation}\label{eq:Ep}
  E_\text{p}(T) = \tfrac12 G \left[
    n \left( \frac n 4 - \Omega - \frac 3 2 \right) + T(T+1) \right] ,
\end{equation}
where $4\Omega$ is the valence space dimension~\cite{ref:Ed52}. We
notice in passing that the fact that the expression \eqref{eq:Ep}
cannot be written in the form of the expression \eqref{eq:Em} entails
that in the lowest-energy eigenstates of $v_\text{p}$, the centroid
\eqref{eq:rho-centr} varies with $T$. In the pure monopole
approximation, every state with the same $T$ has the same energy.
Therefore, if one sets $v = v_\text{m} +v_\text{p}$, one gets just
$E(T) = E_\text{m}(T) + E_\text{p}(T)$. In this approximation,
$- E_\text{p}(T)$ should thus account for the discrepancies
$E_\text{m}(T) - E(T)$. In particular, this difference should not
depend on $J_\text{cut}$ and $J'_\text{cut}$, which is roughly what
was observed for $T = 0$ and 4.

When the $J = 0$ interaction is approximated by $v_\text{p}$, it must
be excluded from the centroids \eqref{eq:set} for consistency. This
amounts to subtracting the last row of results in Table~\ref{tbl:kb3}
from every row. In the following, every reference to results in the
monopole approximation is therefore to the $1f_{7/2}$ results in the
table thus modified. In the monopole plus pairing approximation, the
coefficient of $\tfrac12 T(T+1)$ is $G_0 - G_1 + G$ by
Eqs.~\eqref{eq:Em} and \eqref{eq:Ep}. This is seen to be always
positive when $G = 0.48$~MeV. We now have a correlation between the
variations of $E(0)$ and the width $E(4) - E(0) = 10 (G_0 - G_1 + G)$
of the symmetry spectrum: When an attractive component with $J > 0$ is
turned off, either $G_0$ or $G_1$ decreases and, accordingly, $E(0)$
increases. On the other hand, the effect on $E(4) - E(0)$ depends on
whether the component that is turned off has $T' = 0$ or 1. If it has
$T' = 0$, then $G_0$, and therefore $E(4) - E(0)$, decreases, as well
(as required when $E(0)$ approaches the constant $E(4)$ from below).
If it has $T' = 1$, then $G_1$ decreases, so $E(4) - E(0)$ increases.
The changes are opposite when the component that is turned off is
repulsive. Exactly this pattern is found in Fig.~\ref{fig:ekb3}.

Now consider the case when every interaction but the one in the
$J = 0$ channel has been turned off. In our approximation, the width
of the symmetry spectrum then is
$E_\text{p}(4) - E_\text{p}(0) = 10 \, G = 4.8$~MeV. This fairly
agrees with $E(4) - E(0) = 6.1$~MeV in the KB3 calculation. With
$\Omega = 10$, corresponding to the full \textit{pf} shell,
Eq.~\eqref{eq:Ep} gives $-E_\text{p}(0) = 18.2$~MeV, which is much
larger than $-E(0) = 9.7$~MeV from KB3. This must be due to the spread
of the single-nucleon levels, which makes the higher orbits
participate less effectively in the pair correlations. Taking, on the
other hand, $\Omega = 4$, corresponding to the $1f_{7/2}$ shell alone,
gives $-E_\text{p}(0) = 6.7$~MeV, which is too small. The $J = 0$
component of the KB3 interactions thus causes some scattering out of
the $1f_{7/2}$ shell, so the dominance of the $1f_{7/2}$ shell does
not mean that the nucleons stay there. Despite the scattering out of
it by the $J = 0$ interaction, the contribution to the total energy of
the rest of the interaction seems, however, mainly determined by its
$1f_{7/2}$ part.

On a final note, we point out that by Eqs.~\eqref{eq:Em} and
\eqref{eq:Ep}, the symmetry energy is proportional to $T(T+1)$ in the
monopole plus pairing approximation.

\subsection{\label{sec:24}$A = 24$}

\begin{table}
  \caption{\label{tbl:usd}As Table~\ref{tbl:kb3} for $A = 24$ and the USD
    Hamiltonian. Like in the previous table, results are not shown when
    they do not differ from those in the preceding row because the
    last $T'$ and $J$ turned off is not accommodated by the valence
    space.}
\begin{ruledtabular}
\begin{tabular}{ddddddd}
\multicolumn{1}{c}{$J_\text{cut}$ or}&
\multicolumn{2}{c}{$1d_{5/2}$}&
\multicolumn{2}{c}{$1d_{5/2}+2s_{1/2}$}&
\multicolumn{2}{c}{\textit{sd} shell}\\
\multicolumn{1}{c}{$J'_\text{cut}$}&
\multicolumn{1}{c}{$E_\text{m}(0)$}&
\multicolumn{1}{c}{$G_0\!-\!G_1$}&
\multicolumn{1}{c}{$E_\text{m}(0)$}&
\multicolumn{1}{c}{$G_0\!-\!G_1$}&
\multicolumn{1}{c}{$E_\text{m}(0)$}&
\multicolumn{1}{c}{$G_0\!-\!G_1$}\\\hline
0&-72.22&2.33&-67.55&2.58&-67.30&2.71\\
1&-69.89&2.09&-63.48&2.17&-60.72&2.05\\
2&&&-61.47&1.97&-56.18&1.60\\
3&-64.88&1.59&-51.04&0.93&-48.29&0.81\\
4&&&&&-43.09&0.29\\
5&-42.75&-0.62&-38.13&-0.36&-37.13&-0.31\\
4&-40.98&-0.52&-37.18&-0.31&-33.17&-0.09\\
3&&&-40.61&-0.50&-35.75&-0.23\\
2&-34.97&-0.19&-34.76&-0.18&-32.18&-0.03\\
1&&&&&-33.53&-0.11
\end{tabular}
\end{ruledtabular}
\end{table}

Much of the preceding discussion applies, as well, to the case
$A = 24$, with obvious modifications. To a large extent the orbit
$1d_{5/2}$ replaces $1f_{7/2}$. One difference between the two cases
was touched upon already: Isospin $T = 4$ can be generated in the
\textit{pf} shell with eight nucleons in the $1f_{7/2}$ shell while in
the \textit{sd} shell it requires two nucleons outside the $1d_{5/2}$
shell. These two nucleons can be accommodated by the $2s_{1/2}$ shell,
which is next in energy. Results in the monopole approximation are
displayed in Table~\ref{tbl:usd}. In these calculations,
$\text{tr\,}\rho_1 h = -31.58$~MeV, which is eight times the energy of
the $1d_{5/2}$ orbit in the USD Hamiltonian. Three different $\mathcal
S$ are considered: the $1d_{5/2}$ shell, the $1d_{5/2}$ and $2s_{1/2}$
shells, and the full \textit{sd} shell.

A comparison between Fig.~\ref{fig:eusd} and the results with
$1d_{5/2}$ centroids in Table~\ref{tbl:usd} reveils the same
correlation between the shift of $E(0)$ when a component of the USD
interaction with odd $T' + J$ is turned off and the corresponding
$1d_{5/2}$ shell matrix element as found for the KB3 interaction and
the $1f_{7/2}$ shell. Also, in most instances, like for $A = 48$, the
symmetry spectrum almost does not change when components of the USD
interaction with even $T' + J$ are turned off. This has one
conspicuous exception: $E(0)$ decreases noticeably when the
$T' = 1, J = 3$ component is turned off. The calculation with
$1d_{5/2} + 2s_{1/2}$ centroids gives a hint to an explanation: The
only $T' = 1, J = 3$ interaction in the $1d_{5/2} + 2s_{1/2}$ valence
space, namely the diagonal interaction in the configuration
$1d_{5/2}2s_{1/2}$, is repulsive. This repulsive matrix element
actually produces with $1d_{5/2} + 2s_{1/2}$ centroids a shift that is
much larger than the shift in the USD calculation, namely $-3.43$~MeV
as compared to $-1.25$~MeV. The calculation with $1d_{5/2} + 2s_{1/2}$
centroids also suggests an upwards shift of $E(0)$ by 2.01~MeV when
the $T' = 0, J = 2$ component of the USD interaction is turned off
(due again to interaction in the configuration $1d_{5/2}2s_{1/2}$).
The USD energy indeed shifts upwards by 1.23~MeV. The calculation with
\textit{sd} shell centroids suffers from discrepancies with the USD
results similar to those between the calculation with \textit{pf}
shell centroids and the KB3 results. The present analysis thus
suggests a dominance of the $1d_{5/2} + 2s_{1/2}$ valence subspace in
the USD results.

With $1d_{5/2}$ centroids, we get $E_\text{m}(0) - E(0) = 14.9$~MeV
and $E_\text{m}(4) - E(4) = -7.8$~MeV for $J_\text{cut} = 0$, and
\linebreak $E_\text{m}(0) - E(0) = 10.4$~MeV and
$E_\text{m}(4) - E(4) = -1.1$~MeV for $J'_\text{cut} = 1$. (The
increase of $E_\text{m}(4) - E(4)$ mainly occurs between
$J'_\text{cut} = 4$ and 3, where, as seen from Fig.~\ref{fig:eusd},
$E(4)$ drops even more than $E(0)$ due to the removal of the repulsive
$T' = 1, J = 3$ interaction in the configuration $1d_{5/2}2s_{1/2}$
while $E_\text{m}(4)$ stays constant.) As these changes are relatively
small compared to the total increase of $E(0)$, we conclude, like for
$A = 48$, that with $1d_{5/2}$ centroids, the difference
$E_\text{m}(T) - E(T)$ is roughly independent of $J_\text{cut}$ and
$J'_\text{cut}$ for $T = 0$ and 4.

The USD matrix elements $V_{10}(jjj'j')$ are negative and the ratio of
the largest and smallest $G$ extracted from them by means of
Eq.~\eqref{eq:vp} is 2.1. The value $G = 0.94$~MeV given by the
$1d_{5/2}$ diagonal matrix element is close to the arithmetic and
geometric means. It renders $G_0 - G_1 + G$ positive for all
$G_0 - G_1$ in Table~\ref{tbl:usd} after subtraction of the last row
of results from every row. The pattern of narrowing and widening of
the symmetry spectrum then suggested by the calculation with
$1d_{5/2} + 2s_{1/2}$ centroids materializes in Fig.~\ref{fig:eusd}.
Again there is one exception: When the $T' =1, J = 4$ interaction is
turned off, the spectrum narrows slightly while it is predicted to
widen. We have no explanation of this. It may be related to another
unexplained anomaly for $J'_\text{cut} = 5$ and 4 discussed in
Sec.~\ref{sec:eW}.

The pairing energies $E_\text{p}(0)$ calculated from Eq.~\eqref{eq:Ep}
with $\Omega = 6$, 4, and 3, corresponding to the full $sd$ shell and
the $1d_{5/2} + 2s_{1/2}$ and $1d_{5/2}$ valence spaces are 20.7,
13.2, and 9.4~MeV as compared to an additional binding energy of
13.7~MeV induced by turning on the $J = 0$ component of the USD
interaction. This suggests that mainly the $1d_{5/2}$ and $2s_{1/2}$
shells take part in pair correlations. The resulting width of the
symmetry spectrum is 9.6~MeV as compared to $10 G = 9.4$~MeV.

\section{\label{sec:eW}Term $\epsilon_\text{W}$}

\begin{figure}
  \includegraphics{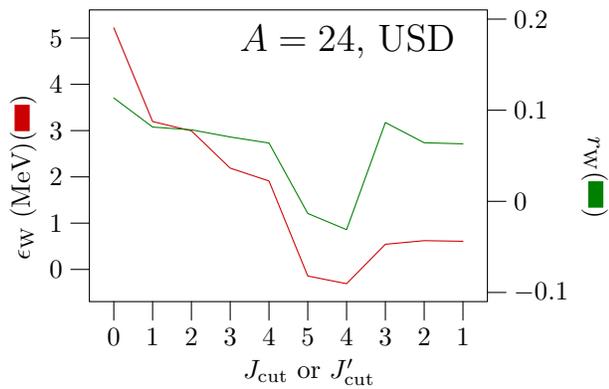}
  \caption{\label{fig:pusd}(Color online) Parameter
    $\epsilon_\text{W}$ and ratio $r_\text{W}$ extracted from the
    energies in Fig.~\ref{fig:eusd}.}
\end{figure}

\begin{figure}
  \includegraphics{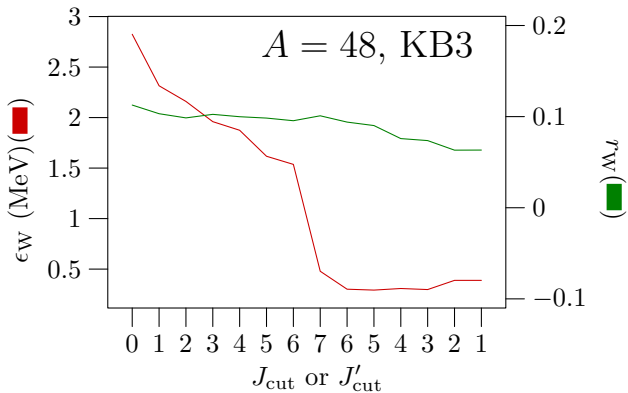}
  \caption{\label{fig:pkb3}(Color online) Parameter
    $\epsilon_\text{W}$ and ratio $r_\text{W}$ extracted from the
    energies in Fig.~\ref{fig:ekb3}.}
\end{figure}

Figures~\ref{fig:pusd} and \ref{fig:pkb3} shows the term
$\epsilon_\text{W}$ in Eq.~\eqref{eq:epWkap} given by the energies in
Figs.~\ref{fig:eusd} and \ref{fig:ekb3}. Readers may verify that the
left hand parts of these plots coincide with those in Fig.~2 of
Ref.~\cite{ref:Sa97} when $\epsilon_{\text{W},J_\text{cut}=0}$ is
normalized to 1. Also shown in Figs.~\ref{fig:pusd} and \ref{fig:pkb3}
is the ratio
\begin{equation}\label{eq:rW}
  r_\text{W} = \frac {\epsilon_\text{W}} { E(4) - E(0) } .
\end{equation}
For both $A$ one gets $r_\text{W} = 0.113$ for $J_\text{cut} = 0$ and
$r_\text{W} = 0.063$ for $J'_\text{cut} = 1$. The latter value is
close to what results when the symmetry energy is proportional to
$T(T+1)$, namely $r_\text{W} = 1/15 = 0.067$. Thus, despite the
single-nucleon spectra being non-degenerate and the $J = 0$
interactions being not exact pairing interactions,
the $J_\text{cut}'= 1$ Hamiltonians are sufficently close to the
pairing Hamiltonian with degenerate single-nucleon levels to almost
maintain the $T(T+1)$ proportionality of the symmetry energy.

There is a one-to-one correspondence between the ratio $r_\text{W}$
and the parameter $x$ in
\begin{equation}\label{eq:thx}
  E(T) = E(0) + \frac {T(T+x)} {2\theta} \quad
  \text{for $T = 2$ and 4} ,
\end{equation}
the precise relation being
\begin{equation}\label{eq:x&rW}
  x = \frac { 12 r_\text{W} } { 1 - 3 r_\text{W} } .
\end{equation}
With $r_\text{W} = 0.113$, this gives $x = 2.05$, which is close to the
values extracted by Bentley and Frauendorf from the measured binding
energies in the $A = 24$ and 48 isobaric chains~\cite{ref:Be13}. This
is natural because the USD and KB3 Hamiltonians were fitted partly to
these binding energies. In the compilation by Bentley and Frauendorf,
$x$ has, as a function of $A$, local maxima at $A = 24$ and 48. These
local maxima result in Nilsson-Strutinskij calculations by
Bentley\etal\ which reasonably reproduce the data, from large
deformations of $^{24}$Mg and $^{48}$Cr~\cite{ref:Be14}. It is not
evident how this translates to interacting nucleons in a single
oscillator shell.

The variation of $r_\text{W}$ from $J_\text{cut} = 0$ to
$J'_\text{cut} = 1$ is markedly different in the two cases of $A = 24$
and 48. For $A = 48$ it is gentle and almost monotonic while for $A =
24$ it proceeds violently from $J_\text{cut} = 4$ to $J'_\text{cut} =
2$. If one neglects, however, the three cases of $J'_\text{cut} = 5$,
4, and 3, the variation for $A = 24$ is similar to the one for $A =
48$. We have no explanation of the apparent anomaly pertaining to
these three $J'_\text{cut}$, but whatever its reason, it is evidently
a non-generalizable feature of $A = 24$ and the USD Hamiltonian. It
may be noticed that $r_\text{W}$ measures the partition of the interval
from $E(0)$ to $E(4)$ by the energy $E(2)$. Thus,
\begin{equation}\label{eq:rW&r2}
  r_\text{W} = \frac {4 r_2 - 1} 3 , \quad
  r_2 = \frac {E(2) - E(0)} {E(4) - E(0)} .
\end{equation}
Because of the factor $4/3$ in this relation, fairly small variations
of $r_2$ result in variations of $r_\text{W}$ which are large in
comparison with the typical $r_\text{W} \approx 0.1$.

In a gross picture, $r_\text{W}$ thus varies with $J_\text{cut}$ and
$J'_\text{cut}$ gently within narrow limits. This suggests that the
drastic decrease of $\epsilon_\text{W}$ observed in
Ref.~\cite{ref:Sa97} when the $T' = 1$ interaction is turned off be
seen mainly as a side effect of the equally drastic decrease of $E(4)
- E(0)$ rather than an independent manifestation of the $T' = 0$
interaction.

\section{\label{sec:oth}Other Hamiltonians}

We mention for completeness that we applied the analysis above also to
some more recent Hamiltonians: in the \textit{sd} shell the
Hamiltonians USDA and USDB~\cite{ref:Br06}, and in the \textit{pf}
shell FPD6~\cite{ref:Ri91} and GXPF1~\cite{ref:Ho04}. The results
given by these other Hamiltonians are only marginally different from
those of USD and KB3, and we shall not discuss them.

For $A = 48$, Bentley\etal\ made~\cite{ref:Be14} the analagous
investigation using the $1f_{7/2}$ valence space and the interaction
model~I of Zamick and Robinson~\cite{ref:Za02}. As noted already, in
this valence space, even $J$ is equivalent to $T' = 1$, and odd $J$ to
$T' = 0$. The matrix elements of Zamick and Robinson are relative ones
with the $J = 0$ matrix element normalized to zero. To make the
interaction attractive, Bentley\etal\ subtracted from all the matrix
elements the highest one, which is the $J = 6$ matrix element. By
construction the interaction then is negative semidefinite and the
$J_{cut}' = 7$ and 6 Hamiltonians are identical. Because the model has
only one single-nucleon level, its energy plays no role. We name this
Hamiltonian ZRI.

\begin{figure}
  \includegraphics{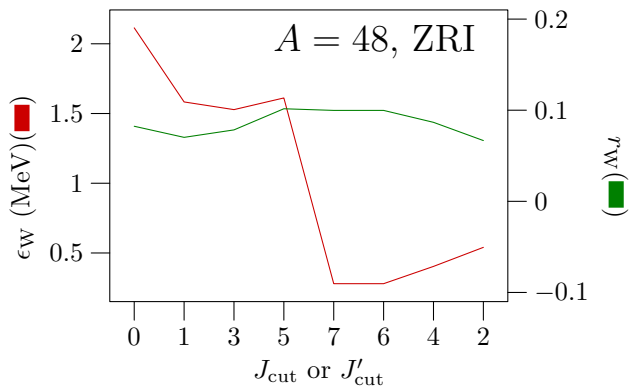}
  \caption{\label{fig:pzri}(Color online)  Parameter
    $\epsilon_\text{W}$ and ratio $r_\text{W}$ extracted from the
    results in Table~I of Ref.~\cite{ref:Be14}.}
\end{figure}

Figure~\ref{fig:pzri} shows the plots similar to Figs.~\ref{fig:pusd}
and \ref{fig:pkb3} derived from the results in Table~I of
Ref.~\cite{ref:Be14}. The ratio $r_\text{W}$ of the full Hamiltonian is
smaller than that of KB3 and therefore also below the empirical value.
Otherwise, the features of the KB3 results discussed above are well
represented by this simpler model. This once again corroborates that
the $1f_{7/2}$ shell seems to dominates these results. As the $J = 0$
part of a single-$j$-shell interaction is an exact pairing
interaction, it gives (with zero $1f_{7/2}$ level) energies in the
form of Eq.~\eqref{eq:Ep}, and $r_\text{W} = 1/15$ in particular.

\section{\label{sec:con}Conclusions}

For mass numbers $A =24$ and 48 we examined with standard isobaric
invariant Hamiltonians for the \textit{sd} and \textit{pf} shells the
variation of the sequence of lowest energies for isospins $T = 0$, 2,
and 4, briefly the symmetry spectrum, when the two-nucleon interaction
is turned off successively in channels with definite isospin and
angular momentum. By definition, these energies form the symmetry
spectrum, and the energy in excess of the $T = 0$ energy is the
symmetry energy. When only the angular momentum $J = 0$ component of
the two-nucleon interaction remains, the symmetry spectrum is close to
that of the pairing Hamiltonian with degenerate single-nucleon
energies, whose symmetry energy is proportional to $T(T+1)$. Turning
on components with $J > 0$ and negative diagonal matrix element in the
$1d_{5/2}^{\,2}$ or $1f_{7/2}^{\,2}$ configuration decreases the
$T = 0$ energy. It widens the symmetry spectrum when the component
that is turned off has isospin $T' = 0$, and narrows it when $T' = 1$.
The effects are opposite if the matrix element is positive. This
behavior was explained in terms of schematic approximations. A
modification for $A = 24$ of this general behavior caused by the
interaction in the $1d_{5/2}2s_{1/2}$ configurations was discussed.

By and large, the ratio $r_\text{W}$ of the term $\epsilon_\text{W}$
in Eq.~\eqref{eq:epWkap}, which was taken in Ref.~\cite{ref:Sa97} as a
measure of the so-called Wigner term in mass formulas, to the width of
the symmetry spectrum varies gently within narrow limits when
components of the two-nucleon are turned off. This suggests that the
decrease of $\epsilon_\text{W}$ observed in Ref.~\cite{ref:Sa97} when
the $T' = 0$ interactions are turned off should be seen mainly as a
side effect of a narrowing of the symmetry spectrum rather than an
independent manifestation of the $T' = 0$ interaction. This narrowing
is well understood by the aforesaid analysis. An extraordinary violent
variation of $r_\text{W}$ when some components are turned off for
$A = 24$ remains unexplained but is certainly anon-generalizable
feature of this particular case.

\begin{acknowledgments}

Arun Kingan was awarded a Richard J. Plano summer 2017 research
internship from the Physics Department of Rutgers University.

\end{acknowledgments}

\bibliography{shellsym4}

\end{document}